\documentclass[copyright,creativecommons]{eptcs}
\providecommand{\event}{SOS 2021} 
\usepackage{breakurl}             
\usepackage{underscore}           

\usepackage{wrapfig}
\usepackage{tikz}
\usepackage{amsmath}
\usepackage{amsfonts}

\newcommand{\Nat}{\mathbb{N}}
\newcommand{\Bool}{\mathbb{B}}
\newcommand{\true}{\textit{true}}
\newcommand{\false}{\textit{false}}
\newcommand{\Real}{\mathbb{R}}
\newcommand{\dist}[2]{\frac{#1}{#2}}
\newcommand{\lsb}{[\![}
\newcommand{\rsb}{]\!]}

\font \aap cmmi10
\newcommand{\at}[1]{\mbox{\aap ,} #1}

\title{Infinite Choice and Probability Distributions\\
       An Open Problem: The Real Hotel}
\author{Jan Friso Groote
\institute{Department of Mathematics and Computer Science\\
Eindhoven University of Technology\\ Eindhoven, Netherlands}
\email{J.F.Groote@tue.nl}}
\def\titlerunning{Infinite Choice and Probability Distributions}
\def\authorrunning{Jan Friso Groote}
\begin{document}
\maketitle

\begin{abstract}
We sketch a process algebra with data and probability distributions. 
This allows to combine two very powerful abstraction mechanisms namely 
non-deterministic choice and probabilities. However, it is not clear how to
define an appropriate semantics for the generalised choice 
over data in combination with probability density functions. 
The real hotel is a puzzle that exemplifies the core of the problem. 
\end{abstract}

\section{Process algebra with nondeterministic choice and data}
Robin Milner can be seen as the founder of process modelling formalisms, now known
as process calculi or process algebras. Well known process algebras are CCS (Calculus
of Communicating Systems, \cite{DBLP:books/sp/Milner80}), CSP (Communicating Sequential Processes, 
\cite{DBLP:books/ph/Hoare85}) and ACP (Algebra of Communicating Processes \cite{BaetenWeijland90}). 
We work in the setting of mCRL2, which is a process algebra based on ACP, extended with data, time and probabilities
\cite{GrooteMousavi14}.

\begin{wrapfigure}{r}{0.3\textwidth}
    \centering
    \begin{tikzpicture}
        \draw (0,2) circle(1mm) node (s0) {};        
        \draw (0,0) circle(1mm) node (s1) {};

        \draw[->] (0,2.3) -- (s0) ;
        \draw[->] (s0) to[out=-90,in=90] node [left] {$a$} (s1) ;
        \draw[->] (s1) to[out=180,in=-135] node [left] {$b$} (s0);
        \draw[->] (s1) to[out=0,in=-45] node [right] {$c$} (s0);
    \end{tikzpicture}
    \caption{The transition system for $X$ where $X{=}a{\cdot}(b{+}c){\cdot}X$}
    \label{fig:transsys}
\end{wrapfigure}
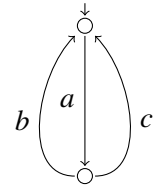
In a contribution in 1973 Robin Milner \cite{Mi73a,Mi73b} pointed out
that the primary concept to model systems with behaviour is the \textit{atomic action},
typically denoted as $a$, $b$, $c$, $\mathit{on}$, $\mathit{off}$. 
Actions are atomic in the sense that they take place at some instant in time. Using process
operators, such as sequential composition (`${\cdot}$') and alternative composition (`$+$'),
actions are combined into behaviour. Recursive behaviour is specified using guarded
recursive equations of the form $X=p$ where $p$ is a process expression. For instance
the process equation $X=a{\cdot}(b+c){\cdot}X$ has as solution for $X$ the process that
can repeatedly do an action $a$, followed
by either an action $b$ or $c$. It is common to depict such behaviour as a transition system,
see Figure \ref{fig:transsys}. There is a special process, denoted $\delta$, (or $\bold{0}$ or \textit{NIL}
in other process algebras), which represents the deadlocked or inactive process that cannot do anything.
By using the parallel operator (`$\parallel$') processes
can be put in parallel, but we do not consider the parallel operator here. 

Milner stressed the importance of \textit{nondeterminism} as an abstract means to 
model complex behaviour in a simple way. A state is nondeterministic if there are multiple
outgoing transitions with the same label to different processes. An example is $a{\cdot}b+a{\cdot}c$.
Whenever the action $a$ happens, it is unclear whether the action $b$ or the action $c$ can be done
next. In the real system that we are describing it may be very clear when which action $a$ is chosen. 
For instance, the first $a$ 
is chosen when the temperature is above 20$^{\circ}$, and otherwise the second $a$ is performed. 
However, it might be that we do not want to model the physics behind temperatures, and simply
abstract away from it. Sometimes people refer to this nondeterminism by `Milner's weather conditions'.

Probabilities provide another abstraction technique and they are very different from nondeterminism. 
When writing $X=(a{\cdot}b+a{\cdot}c){\cdot}X$, the process $X$ repeatedly
performs the nondeterministic $a$. There is some unknown mechanism that determines which $a$ is 
chosen. It could be that they are chosen with a certain probability, but it is also possible
that in every odd round the left $a$ is chosen, and in every even round the right $a$ is selected. 
Nondeterminism is a more liberal way of modelling than probabilities. We come back on probabilities later.
\vspace{2ex}\\
When modelling realistic systems, it is necessary to be able to use data. Here, the four ways
data is added to processes in mCRL2 are explained. Data is specified using applicative equational 
datatypes \cite{GrooteMousavi14}. The details are not relevant for us at the moment,
although the relationship between abstract datatypes and measurable functions requires 
further clarification.
For the moment it is enough to know that all relevant data types such as natural numbers, integers, reals, lists,
messages, functions, etc.\ can be specified easily.

The first two ways of extending processes with data is by using data in actions and  
in process equations. Actions can be extended with data as arguments, as in $a(3,[])$, which is the
action $a$ with value $3$ and an empty list. Actions with arguments behave as ordinary atomic actions,
and the only difference is in essence the extended name. Process variables can be
parameterised with one or more parameters, 
for instance as in $\mathit{Counter}(n{:}\Nat)=\mathit{up}{\cdot}\mathit{Counter}(n{+}1)$,
which models a simple counter. Strictly spoken, the variable $\mathit{Counter}$ is a higher order variable
ranging over functions from natural numbers to processes. 

If process variables have parameters, behaviour should be controllable by the provided data. For this
the if-then-else construct is used, which we denote as $c{\rightarrow}p{\diamond}q$. The idea is that if
condition $c$ is true, the  behaviour of $p$ is executed, and otherwise the behaviour of $q$. 

Receiving data from some external party is modelled by writing all possible actions representing
the receipt of some data. Concretely, reading either true or false can be denoted by 
\[\mathit{read}(\true){\cdot}X(\true)+\mathit{read}(\false){\cdot}X(\false)\]
where $X(b{:}\Bool)$ represents some process that expresses what will be done with the received boolean.
The outside world indicates whether $\mathit{read}(\true)$ or $\mathit{read}(\false)$ will take place, 
indicating which boolean is the input. 

This can be generalised to data types with an infinite number of elements, such as $\Nat$ or $\Real$,
for which we introduce the generalised sum operator $\sum_{d{:}D}p$. The process expression
\[ \sum_{r:\Real}\mathit{read}(r){\cdot}Y(r)   \]
represents the uncountably infinite sum over real numbers. External processes can indicate which $\mathit{read}(r')$ must
take place, indicating that this $r'$ is the input. Note that the generalised sum operator is a fundamental
operator and not an abbreviation. It acts as a binder for the variable $r$. 

The combination of the generalised sum and the if-then-else construct is very expressive,
both theoretically  \cite{Luttik03} and practically. It is for instance easily possible to
express that a number smaller than 100 must be read and processed:
\[\sum_{n{:}\Nat}(n{<}100)\rightarrow \mathit{read}(n){\cdot}\mathit{Process}(n)\diamond\delta.\]

Although the language presented above is concise, it is more than adequate to model realistic behaviour.
It forms the essence of the mCRL2 process
specification language. This language has been used to model and study many protocols and distributed algorithms.
It is for instance also in use as the verification backend of design tools for high quality control software
\cite{DBLP:conf/fm/Broadfoot05,Dezyne2021}.

\section{Process algebra with probability distributions}
An interesting question is how to extend this process language with the possibility to express probabilities and reason
about them. There are a number of languages available where processes have been combined with probabilities, such 
as Modest \cite{DBLP:conf/tacas/HartmannsH14} and Prism \cite{DBLP:conf/cav/KwiatkowskaNP11}. We follow the approach as outlined in \cite{GrooteLanik2011}.
This approach is partly implemented in the mCRL2 toolset, where state space exploration and reduction for finite
probability distributions is possible.

The essential extension is to write \[\dist{f(d)}{d{:}D}X(d)\]
expressing that a value from a domain $D$ is chosen with probability distribution $f$ and this value is
assigned to variable $d$. This is a concise but flexible 
notation both allowing for discrete and continuous distributions, not restrained to for instance
only exponential distributions as is not uncommon in other probabilistic process algebras. The distribution
$f$ has the property that the sum of $f(d)$ over all elements $d{:}D$ equals $1$. 
The notation in the mCRL2 toolset for this construct is $\textbf{dist}~d{:}D[f(d)].X(d)$.

It is pretty straightforward to describe probabilistic behaviour. We give a number of simple examples.
The process throwing a head or tail repeatedly and with equal probability can be characterised by the equation:
\begin{equation}
\label{eq:coin} 
\mathit{Throw}= (\dist{\frac{1}{2}}{b{:}\Bool}~ b{\rightarrow} \mathit{head}\diamond\mathit{tail}){\cdot}\mathit{Throw}.
\end{equation}

One may wonder whether this process is behaviourally equivalent to a process where it is a priori determined
that head
will be thrown $k\in\Nat$ times, before tail shows up, where the probability of a sequence of $k$ heads is 
$\frac{1}{2}^{k+1}$. This latter behaviour is described by the following two process equations:
\[
\begin{array}{l}
\mathit{ThrowSequence}=\dist{\frac{1}{2}^{k+1}}{k{:}\Nat} \mathit{Heads}(k),\\
\mathit{Heads}(k{:}\Nat)=(k>0){\rightarrow} \mathit{head}{\cdot}\mathit{Heads}(k{-}1)\diamond\mathit{tail}{\cdot}\mathit{ThrowSequence}.
\end{array} \]

An example taken from \cite{GrooteLanik2011} shows how timed behaviour can be expressed.
\[
\mathit{install}\at\mathit{st}{\cdot}\dist{\mathcal{N}_0^{\infty}(\mu,\sigma^2)}{t{:}\Real}\mathit{break\_down}\at{(\mathit{st}+t)}{\cdot} \dist{\mathcal{N}_0^{\infty}(\mu,\sigma^2)}{t{:}\Real}\mathit{repair}\at(\mathit{st}{+}t{+}u)
\]
where $t$ and $u$ are distributed according to the normal distribution $\mathcal{N}^{\infty}_0(\mu,\sigma^2)$ 
truncated to the interval $[0,\infty)$ and $a\at{t}$ expresses that action $a$ happens at exactly time $t$. In the 
mCRL2 tool this is expressed as $a@t$. 

This formalism is also suitable for more complex probabilistic processes. 
For instance in \cite{DBLP:journals/siamrev/GrooteWZ16} a precise behavioural description
is provided of a game of chance called `The game of the Goose', including an analysis of probabilities of the various geese
to win the game. 
In \cite{DBLP:conf/birthday/GrooteV17} an intriguing puzzle called the `Lost boarding pass' is modelled.
Tools can easily provide the requested probability that the last passenger that enters the plane will sit
on his own seat, even for planes with hundreds of thousands of passengers. 

\section{Semantics of probabilistic processes based on measurable sets}
In \cite{GrooteLanik2011} the semantics of probabilistic processes has been described but the 
generalised sum operator has been omitted. In order to understand the semantics of probabilistic
processes, we first explain a common approach to probabilistic processes with distributions
over finite domains.

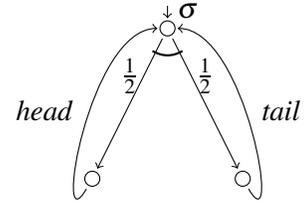
\begin{wrapfigure}{r}{0.3\textwidth}
    \centering
    \begin{tikzpicture}
        \draw (0,2) circle(1mm) node (s0) {} node [above right] {$\sigma$};        
        \draw (-1,0) circle(1mm) node (s1) {};        
        \draw (1,0) circle(1mm) node (s2) {};

        \draw[->] (0,2.3) -- (s0) ;
        \draw[->] (s0) to node [above] {$\frac{1}{2}$} (s1);
        \draw[->] (s0) to node [above] {$\frac{1}{2}$} (s2);
        \draw[thick] (-0.2,1.7) to [out=-30,in=-150] (0.2,1.7);
        \draw[->] (s1) to[out=-120,in=180] node [above left] {$\mathit{head}$} (s0);
        \draw[->] (s2) to[out=-60,in=0] node [above right] {$\mathit{tail}$} (s0);
    \end{tikzpicture}
    \caption{The transition system representing a fair coin}
    \label{fig:prob}
\end{wrapfigure}

The essential idea is that transitions go from states to distributions. These distributions indicate
the probability to end up in a particular state after doing the transition. Consider the behaviour
of the process $\mathit{Throw}$ defined
in process equation (\ref{eq:coin}). It is depicted in Figure \ref{fig:prob}. The behaviour of 
$\mathit{Throw}$ is a distribution $\sigma$ where with probability $\frac{1}{2}$ the behaviour of the left
state, corresponding to doing the action $\mathit{head}$, and with the same probability the behaviour
belonging to the right state is selected. Via the actions $\mathit{head}$ or $\mathit{tail}$
the system ends up in the initial distribution $\sigma$, determining which action will take place next.
Note that if the number of states is finite, 
it is fine if $\sigma$ is a function that for each state provides the probability
that that particular state is taken. 

Now consider the following probabilistic process. 
\[a{\cdot}\dist{\lambda e^{-\lambda r}}{r{:}\Real} ~b(r).\]
First, a single $a$ is done, after which the process ends in an exponential distribution determining the
distribution of values assigned to the variable $r$, which is an argument of the subsequent action $b$. 
Now the probability of ending up in a
state where for some concrete $r' \in \Real$ the action $b(r')$ can be done is of course $0$. As this
is not particularly useful, we provide a density function which provides for sets of states the probability
that one of these states can be reached. For our concrete example the probability of ending up in a state
where an action $b(r'')$ can be done with $r_1<r''<r_2$ is $\int_{r_1}^{r_2} \lambda e^{-\lambda r}d r$.

Abstractly, the distribution $\sigma$ is a measurable function in a sigma algebra. Measurable functions
have a number of pleasant properties, one of them being that the product of two measurable functions is again a
measurable function. This allows us to give semantics to most process operators, in particular the choice operator.
Consider a process $a{\cdot}(p+q)$. Processes $p$ and $q$ induce measurable functions $\sigma_p$ and $\sigma_q$
that tell the probabilities of the various behaviours that can be done by $p$ and $q$. The distribution
of the process $p+q$ is the function $\lambda X{\subseteq}S.\sigma_p(X)\sigma_q(X)$ where $S$ is the set of all states
and $X$ ranges over measurable sets of states.

In \cite{GrooteLanik2011} a semantics for full mCRL2 with distributions is given, including a notion of strong
bisimulation, which is shown to be a congruence. The only operator that was omitted is the generalised
sum operator $\sum_{d:D} p(d)$, with $D$ an infinite, not necessarily countable domain. 
The natural way to define a distribution for this process is to formulate it as the infinite product of all
distributions, aka measurable functions belonging to the processes $p(d)$. Concretely, if $\sigma_d$ is a measurable
function indicating the probability density function for process $p(d)$, the natural definition
for the probability density function for $\sum_{d:D} p(d)$ would be
\[\lambda X{\subseteq} S.\prod_{d:D} \sigma_d(S).\]
It is however unclear what such a product is, or even whether, or under which circumstances, 
it actually exists. But in order to work safely with 
process languages with infinite choice and probability density functions, we require an answer. 

\section{The real hotel}
In order to make the problem of the previous section accessible, we reformulated its essence
as a simply looking puzzle formulated as the \textit{real hotel}.
The puzzle is an analogy of Hilbert's hotel. Hilbert's hotel has a countably infinite number of rooms,
numbered from $0$ upwards. Although the hotel is full, it can still accommodate any additional
countable number of persons, for instance, by moving each guest from room $n$ to room $2n$ freeing up all
odd numbered rooms. 

\begin{wrapfigure}{r}{0.3\textwidth}
    \begin{center}
    \vspace{-4ex}
    \begin{tikzpicture}
        \draw (0,0) rectangle (3,3.3) node (s0) {};        
        \draw (1.5,3.55) node (s1) {\Huge\textit{a}};
        \draw (s1) ++(-0.56,0.56) -- (s1) +(-0.70,0.70);
        \draw (s1) ++(-0.40,0.69) -- (s1) +(-0.5,0.87);
        \draw (s1) ++(-0.21,0.77) -- (s1) +(-0.26,0.97);
        \draw (s1) ++(-0.00,0.80) -- (s1) +(0,1.00);
        \draw (s1) ++(0.56,0.56) -- (s1) +(0.70,0.70);
        \draw (s1) ++(0.40,0.69) -- (s1) +(0.5,0.87);
        \draw (s1) ++(0.21,0.77) -- (s1) +(0.26,0.97);
        \draw (0.15,0.15) rectangle (2.85,0.6) {}; 
        \draw (0.39,0.375) node {\scriptsize $\cdots$};
        \draw (0.6,0.15) rectangle (0.9,0.6) node [pos=0.5] {\tiny $\frac{1}{\sqrt{2}}$}; 
        \draw (1.15,0.375) node {\scriptsize $\cdots$};
        \draw (1.35,0.15) rectangle (1.75,0.6) node [pos=0.5] {\tiny $\frac{1}{\sqrt[3]{2}}$}; 
        \draw (2.2,0.375) node {\scriptsize $\cdots$};
        \draw (2.55,0.15) rectangle (2.85,0.6) node [pos=0.5] {\scriptsize $1$};         
        \draw (0.15,0.75) rectangle (2.85,1.2) node (s0) {}; 
        \draw (0.55,0.975) node {\scriptsize $\cdots$};
        \draw (1.0,0.75) rectangle (1.4,1.2) node [pos=0.5] {\tiny $\frac{1}{\sqrt{\pi}}$}; 
        \draw (1.75,0.975) node {\scriptsize $\cdots$};
        \draw (2.1,0.75) rectangle (2.4,1.2) node [pos=0.5] {\tiny $\frac{1}{\sqrt{3}}$}; 
        \draw (2.65,0.975) node {\scriptsize $\cdots$};        
        \draw (0.15,1.35) rectangle (2.85,1.8) node (s0) {}; 
        \draw (0.34,1.575) node {\scriptsize $\cdots$};
        \draw (0.5,1.35) rectangle (0.8,1.8) node [pos=0.5] {\tiny $\frac{1}{\pi}$}; 
        \draw (1.15,1.575) node {\scriptsize $\cdots$};
        \draw (1.45,1.35) rectangle (1.65,1.8) node [pos=0.5] {\tiny $\frac{1}{e}$}; 
        \draw (1.95,1.575) node {\scriptsize $\cdots$};        
        \draw (2.2,1.35) rectangle (2.5,1.8) node [pos=0.5] {\tiny $\frac{1}{\sqrt{5}}$}; 
        \draw (2.7,1.575) node {\scriptsize $\cdots$};
        
        \draw (0.15,1.95) rectangle (2.85,2.4) node (s0) {}; 
        \draw (0.8,2.175) node {\scriptsize $\cdots$};
        \draw (1.3,1.95) rectangle (1.6,2.4) node [pos=0.5] {\tiny $\frac{1}{\pi^2}$}; 
        \draw (2.3,2.175) node {\scriptsize $\cdots$};
        \draw (0.15,2.55) rectangle (2.85,3.0) node (s0) {}; 
        \draw (0.45,2.775) node {\scriptsize $\cdots$};
        \draw (0.7,2.55) rectangle (1.3,3.0) node [pos=0.5] {\tiny $\frac{1}{\sqrt{10001}}$}; 
        \draw (1.65,2.775) node {\scriptsize $\cdots$};
        \draw (2.0,2.55) rectangle (2.3,3.0) node [pos=0.5] {\tiny $\frac{1}{\pi^3}$}; 
        \draw (2.6,2.775) node {\scriptsize $\cdots$};

    \end{tikzpicture}
    \end{center}
    \caption{The real hotel}
    \label{fig:realhotel}
        \vspace{-4ex}
\end{wrapfigure}
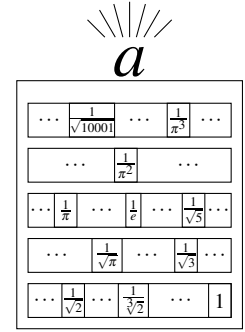
In the real hotel there are an uncountable number of rooms, numbered with a real number ranging
from $0$ up to and including $1$. All rooms are occupied by the participants of a very successful conference. 
In the evening they all arrive at their rooms in the hotel where they find a box from which they
uniformly draw a real number between $0$ and $1$. If the drawn number matches the room number, 
the participant presses a big red button, marked $a$. If one of the participants presses this 
button a big letter {\LARGE \textit{a}} on the roof of the hotel is lit. The question is to determine the probability
that the light on the roof of the hotel will switch on. 

As a process this can be modelled as follows, where for brevity we use $\Real_{01}$ as the set
of real numbers in the interval $(0,1]$ and $U_{01}(s)$ represents the uniform distribution in the interval
$(0,1]$. 
\begin{equation}
\label{eq:realhotel}
\sum_{r{:}\Real_{01}}\dist{U_{01}(s)}{s{:}\Real_{01}} ~(r=s)\rightarrow a\diamond \delta.
\end{equation}
In words, the participant who resides in room $r$ draws a number $s$ and if $s$ equals $r$ the
light on the hotel will switch on, represented by the action $a$. Of course, the probability
that a particular guest presses this button is $0$. But as this experiment is repeated 
uncountably often, it is conceivable that the probability that the light on the roof switches on is larger than $0$. 

Actually, I believe that the behaviour in (\ref{eq:realhotel}) 
is exactly bisimilar to the following simpler process, where the
value of $p$ is $1-\frac{1}{e}\approx 0.63$:
\[ \dist{\textit{if}(b,p,1{-}p)}{b{:}\Bool}~b\rightarrow a \diamond \delta.\]

We can get confidence that $p=1-1/e$ by finitely approximating the hotel. 
Divide the guests in $n$ groups occupying rooms with numbers $(\frac{i}{n},\frac{i{+}1}{n}]$ for 
$i=0,\ldots,n-1$. The probability that a guest in group $i$ draws a number in group $i$ is $1/n$. 
The probability that the light stays off is equal to the probability that no participant draws
a number in his own group. So, the probability that the light goes on is given by the following formula, where we take $n\rightarrow\infty$:
\[ \lim_{n\rightarrow\infty}1-\prod_{i=0}^{n-1} (1-\frac{1}{n})=1-\frac{1}{e}.\]

The question is whether this calculation is indeed appropriate. Asking mathematicians with a background
in probability theory invariably led to the answer that the puzzle of the real hotel cannot be answered
due to the fact that it is not well posed. 
The real challenge is to set up a theory in which the puzzle of the real hotel can be solved convincingly, 
or more in general to set up a theory in which process algebra expressions as in (\ref{eq:realhotel}) have a 
well defined and understood semantical meaning. 

As pointed out in \cite{GrooteLanik2011}, the essential step is to define a density function 
\lsb p\rsb which maps the stochastic variables that occur in $p$ to their probability of occurrence.
For a process $p+q$, this typically is the product of the density functions for $p$ multiplied
by the one for $q$, which is well defined. For the generalised sum operator $\sum_{d{:}D}p(d)$ this
is the potentially unbounded product of the density functions for $p(d)$. 
Only a few papers could be found that
provide leads to an understanding of such unbounded products \cite{Aumann61,OttYorke2005}. 

Ignoring the crucial aspect of being well-defined, it is still intriguing to write down 
examples of probabilistic processes to get an impression of the behaviour that one can 
describe and want to study.

For instance a slot machine has software initialising its random generator. The
random generator $D_t(x)$ determines the payout $x$ but it is influenced by the time $t$.
The process becomes:
\[ \mathit{SlotMachine}=\sum_{t:\Real^+}\textit{wait}(t){\cdot} \dist{D_t(x)}{x{:}\Real}\textit{payout}(x){\cdot}\mathit{SlotMachine}.\]
An interesting question is to determine the maximal amount of profit that can be obtained by optimal 
waiting. 

There are other conceivable combinations of choice and probability. In a store there are 
$k$ employees for some fixed number $k$. The management, not knowing the employees' personalities,
uniformly selects one of the employees. Each employee has its own nondeterministic view on how effective he should
work with an average output between of $l_1$ to $l_2$ outputs per hour. We can describe this 
by 
\[ \dist{\frac{1}{k}}{p{:}\mathit{Person}}\sum_{r{:}\Real_{01}} 
\dist{(l_1+r l_2)e^{-l_1-r l_2 s}}{s{:}\Real}\mathit{production}(p,s).\]
Although we have a clear feeling that this expression is natural, the core question is how to make sure
that it has a well defined meaning. 

\section{Conclusion}
We outlined a process algebra with a generalised sum operator and probability distributions.
We argued that the semantics of the combination of these two very important modelling notions is not
properly understood at the moment. This lack of understanding hampers progress in the field 
of behaviour with continuous distributions. When the semantics is properly understood, 
we may open up a whole new field with probabilistic behavioural equivalences, axioms, rules of 
calculations and verification tools. But most importantly, we will be able to describe and
analyse systems with uncountable nondeterminism and probabilities. 

\paragraph{Acknowledgements.} I like to thank Jan Lanik, Pedro D'Argenio and Pedro S\'{a}nchez Terraf
for the useful discussions we had regarding this topic.

\bibliographystyle{eptcsstyle/eptcs}
\bibliography{references}
\end{document}